\documentclass[12pt]{article}
\usepackage{graphicx}
\usepackage{lscape}
\usepackage{amssymb}
\usepackage{appendix}
\usepackage{multirow}

\begin{document}
\def\qq{\langle \bar q q \rangle}
\def\uu{\langle \bar u u \rangle}
\def\dd{\langle \bar d d \rangle}
\def\sp{\langle \bar s s \rangle}
\def\GG{\langle g_s^2 G^2 \rangle}
\def\Tr{\mbox{Tr}}
\def\figt#1#2#3{
        \begin{figure}
        $\left. \right.$
        \vspace*{-2cm}
        \begin{center}
        \includegraphics[width=10cm]{#1}
        \end{center}
        \vspace*{-0.2cm}
        \caption{#3}
        \label{#2}
        \end{figure}
	}
	
\def\figb#1#2#3{
        \begin{figure}
        $\left. \right.$
        \vspace*{-1cm}
        \begin{center}
        \includegraphics[width=10cm]{#1}
        \end{center}
        \vspace*{-0.2cm}
        \caption{#3}
        \label{#2}
        \end{figure}
                }

\def\ds{\displaystyle}
\def\beq{\begin{equation}}
\def\eeq{\end{equation}}
\def\bea{\begin{eqnarray}}
\def\eea{\end{eqnarray}}
\def\beeq{\begin{eqnarray}}
\def\eeeq{\end{eqnarray}}
\def\ve{\vert}
\def\vel{\left|}
\def\ver{\right|}
\def\nnb{\nonumber}
\def\ga{\left(}
\def\dr{\right)}
\def\aga{\left\{}
\def\adr{\right\}}
\def\lla{\left<}
\def\rra{\right>}
\def\rar{\rightarrow}
\def\lrar{\leftrightarrow}  
\def\nnb{\nonumber}
\def\la{\langle}
\def\ra{\rangle}
\def\ba{\begin{array}}
\def\ea{\end{array}}
\def\tr{\mbox{Tr}}
\def\ssp{{\Sigma^{*+}}}
\def\sso{{\Sigma^{*0}}}
\def\ssm{{\Sigma^{*-}}}
\def\xis0{{\Xi^{*0}}}
\def\xism{{\Xi^{*-}}}
\def\qs{\la \bar s s \ra}
\def\qu{\la \bar u u \ra}
\def\qd{\la \bar d d \ra}
\def\qq{\la \bar q q \ra}
\def\gGgG{\la g^2 G^2 \ra}
\def\q{\gamma_5 \not\!q}
\def\x{\gamma_5 \not\!x}
\def\g5{\gamma_5}
\def\sb{S_Q^{cf}}
\def\sd{S_d^{be}}
\def\su{S_u^{ad}}
\def\sbp{{S}_Q^{'cf}}
\def\sdp{{S}_d^{'be}}
\def\sup{{S}_u^{'ad}}
\def\ssp{{S}_s^{'??}}

\def\sig{\sigma_{\mu \nu} \gamma_5 p^\mu q^\nu}
\def\fo{f_0(\frac{s_0}{M^2})}
\def\ffi{f_1(\frac{s_0}{M^2})}
\def\fii{f_2(\frac{s_0}{M^2})}
\def\O{{\cal O}}
\def\sl{{\Sigma^0 \Lambda}}
\def\es{\!\!\! &=& \!\!\!}
\def\ap{\!\!\! &\approx& \!\!\!}
\def\ar{&+& \!\!\!}
\def\ek{&-& \!\!\!}
\def\kek{\!\!\!&-& \!\!\!}
\def\cp{&\times& \!\!\!}
\def\se{\!\!\! &\simeq& \!\!\!}
\def\eqv{&\equiv& \!\!\!}
\def\kpm{&\pm& \!\!\!}
\def\kmp{&\mp& \!\!\!}
\def\mcdot{\!\cdot\!}
\def\erar{&\rightarrow&}


\def\simlt{\stackrel{<}{{}_\sim}}
\def\simgt{\stackrel{>}{{}_\sim}}


\renewcommand{\textfraction}{0.2}    
\renewcommand{\topfraction}{0.8}   

\renewcommand{\bottomfraction}{0.4}   
\renewcommand{\floatpagefraction}{0.8}
\newcommand\mysection{\setcounter{equation}{0}\section}

\def\baeq{\begin{appeq}}     \def\eaeq{\end{appeq}}  
\def\baeeq{\begin{appeeq}}   \def\eaeeq{\end{appeeq}}
\newenvironment{appeq}{\beq}{\eeq}   
\newenvironment{appeeq}{\beeq}{\eeeq}
\def\bAPP#1#2{
 \markright{APPENDIX #1}
\addcontentsline{toc}{section}{Appendix #1 #2}
 \medskip
 \medskip
\begin{center}      {\bf\LARGE Appendix #1 }{\quad\Large\bf #2}
\end{center}
 \renewcommand{\thesection}{#1.\arabic{section}}
\setcounter{equation}{0}
        \renewcommand{\thehran}{#1.\arabic{hran}}
\renewenvironment{appeq}
  {  \renewcommand{\theequation}{#1.\arabic{equation}}
     \beq
  }{\eeq}
\renewenvironment{appeeq}
  {  \renewcommand{\theequation}{#1.\arabic{equation}}
     \beeq
  }{\eeeq}
\nopagebreak \noindent}

\def\eAPP{\renewcommand{\thehran}{\thesection.\arabic{hran}}}

\renewcommand{\theequation}{\arabic{equation}}
\newcounter{hran}
\renewcommand{\thehran}{\thesection.\arabic{hran}}

\def\bmini{\setcounter{hran}{\value{equation}}
\refstepcounter{hran}\setcounter{equation}{0}
\renewcommand{\theequation}{\thehran\alph{equation}}\begin{eqnarray}}
\def\bminiG#1{\setcounter{hran}{\value{equation}}
\refstepcounter{hran}\setcounter{equation}{-1}
\renewcommand{\theequation}{\thehran\alph{equation}}
\refstepcounter{equation}\label{#1}\begin{eqnarray}}


\newskip\humongous \humongous=0pt plus 1000pt minus 1000pt
\def\caja{\mathsurround=0pt}


\title{
         {\Large
                 {\bf
Analysis of heavy spin--3/2 baryon--heavy spin--1/2 baryon--light 
vector meson vertices in QCD
                 }
         }
      }

\author{\vspace{1cm}\\
{\small T. M. Aliev$^a$ \thanks
{e-mail: taliev@metu.edu.tr}~\footnote{permanent address:Institute
of Physics,Baku,Azerbaijan}\,\,,
K. Azizi$^b$ \thanks
{e-mail: kazizi@dogus.edu.tr}\,\,,
M. Savc{\i}$^a$ \thanks
{e-mail: savci@metu.edu.tr} \,\,,
V. S. Zamiralov$^c$ \thanks  
{e-mail: zamir@depni.sinp.msu.ru}} \\
{\small (a) Physics Department, Middle East Technical University,
06531 Ankara, Turkey} \\
{\small (b) Physics Division,  Faculty of Arts and Sciences,
Do\u gu\c s University,} \\
{\small Ac{\i}badem-Kad{\i}k\"oy,  34722 Istanbul, Turkey} \\
{\small (c) Institute of Nuclear Physics, M. V. Lomonosov MSU, Moscow,
Russia} }

\date{}

\begin{titlepage}
\maketitle
\thispagestyle{empty}

\begin{abstract}
The heavy spin--3/2 baryon--heavy spin--1/2 baryon vertices
with light vector mesons are studied
within the light cone QCD sum rules method.
These vertices are parametrized in terms of three coupling constants.
These couplings are calculated for all possible transitions.
It is shown that correlation functions for these transitions are 
described by only one invariant function for every Lorenz structure.
The obtained relations between the correlation functions 
of the different transitions are structure independent while 
explicit expressions of invariant functions depend on the Lorenz 
structure.


\end{abstract}
\end{titlepage}
\section{Introduction}
 In the last decade, significant experimental progress has been achieved in heavy baryon physics.
These highly excited (and often unexpected) experimental results have been announced by the
BaBar, Belle, CDF and D$\emptyset$ Collaborations. 
The ${1\over 2}^+$ and 
${1\over 2}^-$ antitriplet states, $\Lambda_c^{+},~\Xi_c^{+},~\Xi_c^{0}$ 
and $\Lambda_c^{+} (2593)$,
$\Xi_c^{+}(2790),~\Xi_c^{0}(2790)$ as well as the ${1\over 2}^+$
and  ${3\over 2}^+$  sextet states, $\Omega_c,\Sigma_c,\Xi'_c$ and
$\Omega_c^\ast,\Sigma_c^\ast,\Xi_c^\ast$ have been observed 
\cite{RZ01,Rstp01}. Among the S--wave bottom baryons, only
$\Lambda_b,~\Sigma_b,~\Sigma_b^\ast,~\Xi_b$ and $\Omega_b$ have been discovered.
Moreover, the physics at LHC opens new horizons for detailed study 
of the observed heavy baryons, and paves the way to an invaluable opportunity
for search of
the new baryon states \cite{RZ02}. Evaluation of the strong heavy baryon--heavy baryon--meson coupling constants could be important in analysis of
 $\bar{p}p$ and $e^+e^-$ experiments with pair production of heavy baryons with spin--1/2 or spin--3/2. For example, in BaBar and BELLE, 
these baryons can be  produced in pairs, where one of them is off-shell. Analysis of the subsequent of the  strong decays of this baryons requires knowledge about 
their strong   coupling constants.

Considerable progress in experiments has stimulated theoretical analysis of the heavy flavor physics.
Heavy baryons with a single heavy quark can serve as an excellent ``laboratory" for testing predictions
of the quark models and heavy quark symmetry.
After discovery of heavy baryons with a single heavy quark the next step of investigations
in this direction is to study their strong, electromagnetic and weak decays which can give us useful
information on the quark structure of these baryons.

The strong coupling constants of these baryons with light vector mesons are
the main ingredients for
their strong decays and more accurate determination of these constants is
needed. For this aim one should consult to some kind of nonperturbative methods in
QCD as we deal at hadronic scale. The method of the QCD sum rules \cite{RZ03} proved to
be one of the most predictive  among all other nonperturbative methods,
and in this respect,
the most advanced version seems to be the formalism implemented on light cone.
In the light cone QCD sum rules (LCSR), the  operator product expansion (OPE) is carried out near the light cone, $x^2 \sim 0$, and
the nonperturbative hadronic matrix elements are parametrized in terms of distribution
amplitudes (DA's) of a given particle (for more about LCSR, see \cite{RZ04}).

In \cite{RZ05prime}, \cite{RZ05} and \cite{RZ06}, the  strong coupling constants of 
light pseudoscalar and vector mesons with sextet and antitriplet of the
spin--1/2 heavy baryons as well as the heavy spin--3/2 baryon--heavy spin--1/2 baryon vertices
with light pseudoscalar mesons  are calculated within light cone version of the QCD sum
rules.
In the present work, we extend our previous studies to  investigate the
strong coupling constants among sextet of the heavy spin--3/2 baryons  and
the sextet and antitriplet of the heavy spin--1/2 baryons and the light vector mesons.

The plan of this work is as follows. We first derive LCSR for the coupling constants of the
transitions of the sextet spin--3/2 heavy baryons to sextet and antitriplet
spin--1/2 heavy baryons and light vector mesons. In section 3, we present our numerical analysis of the
aforementioned coupling constants and compare our predictions with the
results available on this subject.
\section{Light cone QCD sum rules for $B_Q^*B_QV$ vertices}

In this section, we calculate the strong coupling constants $B_Q^*B_Q^6 V$ and 
$B_Q^*B_Q^3 V$,
where $B_Q^*$ is the heavy spin--3/2 sextet, $B_Q^6$ stands for the heavy spin--1/2 sextet and $B_Q^3$ denotes the
heavy spin--1/2 antitriplet baryons.
The vertex describing spin--3/2 baryon transition into spin--1/2 baryon and light vector meson can be parametrized
in the following way \cite{RZ07}: 
\bea
\label{eZ01}
\lla B_Q(p_2) V(q)| B_Q^*(p_1) \rra \es \bar{u}_{B_Q}(p_2)
\Big\{ g_1 (q_\alpha\
\rlap/{\varepsilon} - \varepsilon_\alpha \rlap/{q} ) \gamma_5 + 
g_2 [(P \cdot \varepsilon) q_\alpha - (P\cdot q) \varepsilon_\alpha] 
\gamma_5 \nnb \\
\ar g_3 [(q\cdot\varepsilon) q_\alpha - q^2 \varepsilon_\alpha] \gamma_5
\Big\}u_{B_{Q\alpha}^* }(p_1),
\eea
where ${u}_{B_Q}(p_2)$ is the Dirac spinor of either the sextet baryon $B^6_Q$ or  the
antitriplet baryon $B^3_Q$, while $u_{B_{Q\alpha}^* }(p_1)$ is the
Rarita--Schwinger spinor of the spin--3/2 sextet baryon $B_Q^*$,
$\varepsilon_\mu$ is the polarization 4-vector of the light vector meson and
$2P=p_1+p_2$, $q=p_1-p_2$. Furthermore, the conditions 
$q^2=m_V^2$ and $(q\varepsilon)=0$ are imposed for
the on-shell vector meson. Here we would like to make the following remark. Many of considered
transitions in this work are  kinematically forbidden. In other words, one of
particles should be off mass shell and therefore "coupling constants"have $q^2$ dependence.  We calculate these  form factors at $q^2=m_V^2$ and assume 
that  in going $q^2$ from this point to   $q_{max}^2$ (in our case $q^2_{max}=(m_1-m_2)^2$ indeed is small), the coupling constants do  not change considerably (for a detailed discussion see \cite{Reinders}).

In order to calculate the strong coupling constants $g_k$, $k=1,2,3$, in the framework of the
LCSR, we start by considering the correlation function:
\bea
\label{eZ02}
\Pi_\mu^{B_Q^*\rar B_Q V} = i \int d^4x e^{ip_2x} \lla V(q) \vel T\Big\{
\eta^{B_Q} (x) \bar{\eta}_\mu^{B_Q^*} (0) \Big\} \ver 0 \rra~,
\eea
where  $\eta_\mu$ and $\eta$ are interpolating 
currents of the $B_Q^*$ or $B^{6,3}_Q$ baryons, respectively, with $T$ being the time ordering operator.

The general form of the interpolating current of the spin--1/2 heavy sextet $B_b^6$
and antitriplet $B_b^3$ baryons can be written in
the following form \cite{RZ09}:
\bea
\label{eZ03}
\eta^{(6)}_Q (q_1,q_2,Q)\es - \sqrt{1\over 2} \epsilon^{abc} \Big[
\Big( q_1^{aT} C Q^b \Big) \gamma_5 q_2^c - \Big( Q^{aT} C q_2^b \Big)
\gamma_5 q_1^c +
\beta \Big( q_1^{aT} C \gamma_5 Q^b \Big) q_2^c \nnb \\
\ek \beta \Big( Q^{aT} C \gamma_5 q_2^b \Big) q_1^c\Big]~, \nnb \\
\eta^{(3)}_Q (q_1,q_2,Q)\es   \sqrt{1\over 6} \epsilon^{abc} \Big[
2 \Big( q_1^{aT} C q_2^b \Big) \gamma_5 Q^c + \Big( q_1^{aT} C Q^b \Big)
\gamma_5 q_2^c + \Big( Q^{aT} C q_2^b \Big)\gamma_5 q_1^c \nnb \\
\ar 2 \beta \Big( q_1^{aT} C \gamma_5 q_2^b \Big) Q^c + 
\beta \Big( q_1^{aT} C \gamma_5 Q^b \Big) q_2^c +
\beta \Big( Q^{aT} C \gamma_5 q_2^b \Big) q_1^c\Big]~, 
\eea
where $a,b,c$ are the color indices, $C$ is the charge conjugation operator,
$\beta$ is an arbitrary parameter and $\beta=-1$ corresponds to
the choice for the Ioffe current \cite{RZ10}.
The interpolating current for the spin--3/2 sextet baryons
can be written as \cite{RZ09}
\bea
\label{eZ04}
\eta_\mu = A \epsilon^{abc} \Big[ \Big( q_1^{aT} C \gamma_\mu q_2^b \Big)
Q^c + \Big( q_2^{aT} C \gamma_\mu Q^b \Big) q_1^c +
\Big( Q^{aT} C \gamma_\mu q_1^b \Big) q_2^c \Big]~.
\eea
 The light quark contents of both sextet and antitriplet heavy spin--1/2 baryons are shown in Table 1.
The values of normalization constant $A$ and the quark flavors $q_1$, $q_2$ and $Q$ for each
member of the heavy spin--3/2 baryon sextet are also  listed in Table 2.


\begin{table}[thb]

\renewcommand{\arraystretch}{1.3}
\addtolength{\arraycolsep}{-0.5pt}
\small
$$
\begin{array}{|l|c|c|c|c|c|c|c|c|c|}
\hline \hline
  & \Sigma_{b(c)}^{+(++)} & \Sigma_{b(c)}^{0(+)}  & \Sigma_{b(c)}^{-(0)} 
  & \Xi_{b(c)}^{-(0)'}    & \Xi_{b(c)}^{0(+)'}
  & \Omega_{b(c)}^{-(0)}  & \Lambda_{b(c)}^{0(+)}
  & \Xi_{b(c)}^{-(0)}     & \Xi_{b(c)}^{0(+)}    \\  \hline
q_1 & u & u & d & d & u & s & u & d & u \\
q_2 & u & d & d & s & s & s & d & s & s \\
\hline \hline
\end{array}
$$
\caption{The light quark content $q_1$ and $q_2$ for the sextet and
anti--triplet baryons with spin--1/2}
\renewcommand{\arraystretch}{1}
\addtolength{\arraycolsep}{-1.0pt}
\end{table}
  


\begin{table}[thb]

\renewcommand{\arraystretch}{1.3}
\addtolength{\arraycolsep}{-0.5pt}
\small
$$
\begin{array}{|l|c|c|c|c|c|c|}
\hline \hline
 & \Sigma_{b(c)}^{*+(++)} & \Sigma_{b(c)}^{*0(+)} & \Sigma_{b(c)}^{*-(0)}  
 & \Xi_{b(c)}^{*0(+)}    & \Xi_{b(c)}^{*-(0)} 
 & \Omega_{b(c)}^{*-(0)}          \\  \hline
 q_1 & u & u & d & u & d & s \\
 q_2 & u & d & d & s & s & s  \\
 A   & \sqrt{1/3} & \sqrt{2/3} & \sqrt{1/3}
     & \sqrt{2/3} & \sqrt{2/3} & \sqrt{1/3} \\
\hline \hline
\end{array}
$$
\caption{The light quark content $q_1$, $q_2$ and normalization constant $A$ for the sextet baryons with
spin--3/2}
\renewcommand{\arraystretch}{1}
\addtolength{\arraycolsep}{-1.0pt}
\end{table}




Using the quark-hadron duality and inserting a complete set of hadronic
states with the same quantum numbers as the interpolating currents $\eta_\mu$
and $\eta^{6,3}$ into correlation function, one can obtain the representation of
$\Pi_\mu(p,q)$ in terms of hadrons. Isolating the ground state contributions coming
from the heavy baryons in the corresponding channels we get
\bea
\label{eZ05}
\Pi_\mu(p,q) \es
{\lla 0 \vel \eta \ver B_Q(p_2) \rra \lla B_Q(p_2)
V(q) \ve B_Q^*(p_1) \rra \over p_2^2 - m_2^2 } \;
{\lla B_Q^*(p_1) \vel \bar{\eta}_\mu \ver 0 \rra \over p_1^2-m_1^2}  +
\cdots \nnb \\
\es -\frac{\lambda_{B_Q}\lambda_{B_Q^*}(\rlap/{p_2+m_2})}{(p_2^2 - m_2^2 )(p_1^2 - m_1^2 )}
\Big\{ g_1 (q_\alpha\
\rlap/{\varepsilon} - \varepsilon_\alpha \rlap/{q} ) \gamma_5 + 
g_2 [(P\cdot \varepsilon) q_\alpha - (P\cdot q) \varepsilon_\alpha] 
\gamma_5 
\nnb \\
\ar g_3 [(q\cdot\varepsilon) q_\alpha - q^2 \varepsilon_\alpha] \gamma_5 \Big\}(\rlap/{p_1+m_1})\Bigg(g_{\alpha\mu} - {\gamma_\alpha \gamma_\mu \over 3 m_1} -
{2 p_{1\alpha} p_{1\mu} \over 3 m_1^2} + {p_{1\alpha} \gamma_\mu -
p_{1\mu} \gamma_\alpha \over 3 m_1} \Bigg).\nnb \\
\eea
where $m_1=m_{B_Q^*}$, $m_2=m_{B_Q}$, and $\cdots$ represent the
contributions of the higher states and continuum. 
In derivation of Eq. (\ref{eZ05}) we have used the following definitions:
\bea
\label{eZ06}
\lla 0 \vel \eta \ver B_Q(p) \rra \es \lambda_{B_Q} u(p,s)~, \nnb \\
\lla 0 \vel {\eta}_\mu \ver B_Q^*(p) \rra \es \lambda_{B_Q^*} {u}_\mu
(p,s)~,
\eea
and
summation over spins has been performed using the relations,
\bea
\label{eZ07}
\sum_s u(p_2,s) \bar{u}(p_2,s) \es \rlap/{p}_2 + m_2 ~, \nnb \\
\sum_s u_\alpha(p_1,s) \bar{u}_\beta(p_1,s) \es -(\rlap/{p}_1 + m_1)
\Bigg(g_{\alpha\beta} - {\gamma_\alpha \gamma_\beta \over 3 m_1} -
{2 p_{1\alpha} p_{1\beta} \over 3 m_1^2} + {p_{1\alpha} \gamma_\beta -
p_{1\beta} \gamma_\alpha \over 3 m_1} \Bigg) ~.
\eea

Here we would like to make following remark. 
The interpolating current $\eta_\mu$ couples
not only to the $J^P = {3\over2}^+$ states, but also to the $J^P =
{1\over2}^-$ states. The corresponding matrix element of the current $\eta_\mu$ between
vacuum and $J^P = {1\over2}^-$ states can be parametrized as follows
\bea
\label{eZ08}
\langle 0 \vel \eta_\mu \ver \widetilde B_Q (p) \rangle =
\widetilde{\lambda} \Bigg( \gamma_\mu - 4
{p_\mu \over {\widetilde m}} \Bigg) u(p,s)~.
\eea
where tilde means a $J^P={1\over 2}^-$ state, and $\widetilde{\lambda}$ and
$\widetilde{m}$ represent
its residue and mass, respectively. 
Using Eqs. (\ref{eZ05}) and (\ref{eZ08}), we see that 
the structures proportional to $\gamma_\mu$ at the right end and to 
$p_{1\mu}=p_{2\mu}+q_\mu$ receive contributions not only from $J^P=\frac{3}{2}^+ $ states but also 
from the $J^P = {1\over2}^-$ states which should be removed.

Another problem is that not all Lorentz structures are independent. 
We can remove both
problems by ordering the
Dirac matrices in a specific way which guarantees the independence of all
the Lorentz structures as well as the absence of the $J^P = {1\over2}^-$
contributions. In the present work, we choose the ordering of the Dirac matrices 
in the form $\gamma_\mu\rlap/{\varepsilon}\rlap/{q}\rlap/{p}\gamma_5$.
Choosing this ordering and using Eq. (\ref{eZ04}),
for the phenomenological part of the
correlation function we get
\bea
\label{eZ09}
\Pi_\mu \es {\lambda_{B_Q} \lambda_{B_Q^*} \over [m_1 - (p+q)^2]}
{1\over (m_2^2 -p^2) } \Big[g_1 (m_1+m_2) \rlap/{\varepsilon} \rlap/{p}
\gamma_5 q_\mu - g_2 \rlap/{q} \rlap/{p} \gamma_5 (p\cdot\varepsilon) q_\mu
+ g_3 q^2 \rlap/{q} \rlap/{p} \gamma_5 \varepsilon_\mu \nnb \\
\ar \mbox{\rm other structures} \Big]~,
\eea
where we have set $p_2=p$. In order to calculate the coupling constants $ g_1$, $g_2$ and $g_3$ from the QCD sum rules,
we should know the expression for the correlation function from the QCD side.
But  first we try to find relations between various invariant functions which would
simplify the calculations of the constants $ g_1$, $g_2$ and $g_3$
considerably for
different channels.
For this aim we will follow the approach given in \cite{RZ11}-\cite{RZ16}   where main
``construction blocks"
have been derived (see also  \cite{RZ05prime}-\cite{RZ06}). These relations are independent from the choice of the explicit Lorenz
structure and automatically take into account violation of the flavor unitary symmetry.

We will also show that all the transitions of spin--3/2 sextet $B_Q^*$ into the sextet and antitriplet
of the spin--1/2 $B_Q$ are described in terms of only one invariant function for each Lorenz structure.
Firstly, we consider sextet--sextet transitions and as an example we start
with the $\Sigma^{\ast 0}_b \rar \Sigma^0_b \rho^0$ transition.
The invariant correlation function for the $\Sigma^{\ast 0} \rar \Sigma^0_b \rho^0$
transition can be written in the general form as
\bea
\label{eZ10}
\Pi^{\Sigma^{\ast 0}_b \rar \Sigma^0_b \rho^0} = g_{\rho^0 uu} \Pi_1(u,d,b) +
g_{\rho^0 dd} \Pi_1^\prime(u,d,b) + g_{\rho^0 bb} \Pi_2(u,d,b)~.
\eea
The interpolating currents for $\Sigma^{\ast 0}_b$ and $\Sigma^0_b$ are symmetric with respect to
the exchange of the light quarks, then obviously
$\Pi_1^\prime(u,d,b)=\Pi_1(d,u,b)$. Moreover, using Eq. (\ref{eZ04}) one can easily obtain that 
$-\Pi_2(u,d,b)$=$\Pi_1(b,u,d)+\Pi_1(b,d,u)$.
Couplings of quarks to $\rho^0$ meson are obtained from the quark current
\bea
\label{eZ11}
J_\mu^{\rho^0} = \sum_{u,d,s} g_{qq\rho} \bar{q} \gamma_\mu q~, 
\eea
and for the $\rho^0$ meson $g_{\rho^0 uu} = -g_{\rho^0 dd} = 1/\sqrt{2}$ and
similarly for $\omega$ and $\phi$ mesons one get
$g_{\omega uu} = g_{\omega dd} = 1/\sqrt{2}$, $g_{\phi ss} = 1$,
all the other couplings to light mesons being zero.

%
The function $\Pi_1(u,d,b)$ describes
emission of the $\rho^0$ meson from $u$, $d$ and $b$ quarks,
respectively, and is formally defined as
\bea
\label{eZ12}
\Pi_1(u,d,b) \es \lla \bar{u} u \vel \Sigma^{*0}_b \bar{\Sigma}^{\ast 0}_b \ver 0
\rra~.
\eea
Using Eqs. (\ref{eZ10})--(\ref{eZ12}) we get,
\bea
\label{nolabel}
\Pi^{\Sigma^{\ast 0}_b \rar \Sigma^0_b \rho^0} ={1\over \sqrt{2}} \Big[
\Pi_1(u,d,b) - \Pi_1(d,u,b) \Big]~.
\eea
In the isospin symmetry limit, the invariant functions for the $\Sigma^{*0}_b\rightarrow \Sigma_b \rho^0$
and $\Sigma^{*0}\rightarrow \Lambda_b \omega^0$ transitions vanish, as is
expected.
The relations among other invariant functions involving neutral vector mesons $\rho$, $\omega$ and $\phi$
can be obtained in a similar way, and we put these relations into the Appendix.

The relations involving charged $\rho$ mesons  require some care.
Indeed, in the $\Sigma^{\ast 0}_b \rar \Sigma^0_b \rho^0$ transition $u(d)$ quarks from baryons $\Sigma^{\ast 0}_b$ and
$\Sigma^{0}_b$ form $\bar uu (\bar dd)$ state, while $d$ ($u$) and $b$
 quarks are spectators. In the case of charged $\rho^+$ meson $d$ quark from baryons $\Sigma^{0}_b$ and
$u$ quark from baryon $\Sigma^{\ast 0}_b$ form the $(\bar ud)$ state while the remaining $d(u)$ $b$ quarks again
are spectators. 
Therefore it is quite natural to expect that these matrix elements should be proportional. Indeed, explicit calculations 
confirm this expectation and we obtain that
\bea
\label{eZ13}
\Pi^{\Sigma^{\ast +}_b \rar \Sigma^0_b \rho^+} \es \lla \bar{d} \vel \Sigma^0_b
\bar{\Sigma}^{\ast +}_b \ver 0 \rra = -\sqrt{2} \lla \bar{d}d \vel \Sigma^0_b
\bar{\Sigma}^{\ast 0}_b \ver 0 \rra \nnb \\
\es - \sqrt{2} \Pi_1(d,u,b)~.
\eea
Replacing $u \lrar d$ in Eq. (\ref{eZ13}), we get
\bea
\label{eZ14}
\Pi^{\Sigma^{\ast -}_b \rar \Sigma^0_b \rho^-} \es -\sqrt{2} \Pi_1(u,d,b)~.
\eea
It should be noted that the relations between invariant functions involving $\rho$ and $\omega$ mesons can also be  obtained from isotopic symmetry argument.
All other relations among invariant functions involving charged $\rho$, $K^\ast$ and $\phi$
mesons are obtained in a similar way with the proper change of quark symbols and
are presented in the Appendix. 
To describe the sextet--sextet transitions, we need to  calculate the invariant function $\Pi_1$. 
For this aim, the correlation function which describes
the transition $\Sigma^{\ast 0} \rar \Sigma^0 \rho^0$ would serve as a good
candidate.

Up to now we have discussed the sextet--sextet transitions and found that all
these transitions involving light vector mesons are described by a single
universal function. We now proceed discussing the sextet--antitriplet
transitions. Our goal here is to show that these transitions, similar to
the sextet--sextet, are also described with the same invariant function. For
this aim let us consider the $\Sigma^{\ast 0} \rar \Lambda_b \rho^0$ transition.

Similar to Eq. (\ref{eZ10}), this transition can be written as
\bea
\label{eZ15}
\Pi^{\Sigma^{\ast 0} \rar \Lambda \rho^0} = g_{\rho^0 uu}
\widetilde{\Pi}_1(u,d,b) + g_{\rho^0 dd} \widetilde{\Pi}_1^\prime(u,d,b) +
g_{\rho^0 bb} \widetilde{\Pi}_2(u,d,b)~,
\eea
where tilde is used to note the difference of the invariant
function responsible for the sextet--antitriplet transition from the
sextet--sextet transition. 
In order to express the $\widetilde{\Pi}$ in terms of $\Pi$, let us first express
the interpolating current of $\Lambda_b$ in terms of sextet current.
Performing similar calculations as is done in \cite{RZ15}, the following
relation between the two currents can easily be obtained
\bea
\label{eZ16}
\eta^{(6)} (q_1 \lrar Q) - \eta^{(6)} (q_2 \lrar Q) \es \sqrt{3}
\eta^{(3)} (q_1,q_2,Q)~, \nnb \\
\eta^{(6)} (q_1 \lrar Q) + \eta^{(6)} (q_2 \lrar Q) \es -
\eta^{(6)} (q_1,q_2,Q)~.
\eea
Using these relations and Eq. (\ref{eZ10}), we construct the following
auxiliary quantities,
\bea
\label{eZ17}
\Pi^{\Sigma_b^{\ast 0} \rar \Sigma_b (u \lrar b) \rho^0} \es g_{\rho bb}
\Pi_1(b,d,u) + g_{\rho dd} \Pi_1^\prime(b,d,u) + g_{\rho uu}
\Pi_2(b,d,u)~, \\
\label{eZ18}
\Pi^{\Sigma_b^{\ast 0} \rar \Sigma_b (d \lrar b) \rho^0} \es g_{\rho uu}
\Pi_1(u,b,d) + g_{\rho bb} \Pi_1^\prime(u,b,d) + g_{\rho dd}
\Pi_2(u,b,d)~.
\eea

From these expressions, we immediately obtain that
\bea
\label{eZ19}
\sqrt{3} \Pi^{\Sigma_b^{\ast 0} \rar \Lambda_b \rho^0} \es g_{\rho uu} \Big[
\Pi_2 (b,d,u) - \Pi_1 (u,b,d) \Big] + g_{\rho dd} \Big[ 
\Pi_1^\prime (b,d,u) - \Pi_2 (u,b,d) \Big] \nnb \\
\ar  g_{\rho bb} \Big[ \Pi_1(b,d,u) - 
\Pi_1^\prime (u,b,d) \Big]~, \\
\label{eZ20}
- \Pi^{\Sigma_b^{\ast 0} \rar \Sigma_b^0 \rho^0} \es
g_{\rho uu} \Big[ 
\Pi_2 (b,d,u) + \Pi_1 (u,b,d) \Big] + g_{\rho dd} \Big[ 
\Pi_1^\prime (b,d,u) + \Pi_2 (u,b,d) \Big] \nnb \\
\ar  g_{\rho bb} \Big[ \Pi_1(b,d,u)+     
\Pi_1^\prime (u,b,d) \Big] \nnb \\
\es - g_{\rho uu} \Pi_1 (u,d,b) - g_{\rho dd} \Pi_1^\prime (u,d,b) -
g_{\rho bb} \Pi_2 (u,d,b)~,
\eea
where in obtaining the last line, we have used Eq. (\ref{eZ10}).            
From this equation, we immediately get,
\bea
\label{eZ21}
- \Pi_2 (b,d,u) \es  \Pi_1 (u,b,d) +  \Pi_1 (u,d,b)~, \\
\label{eZ22}
- \Pi_2 (u,b,d) \es  \Pi_1^\prime (b,d,u) +  \Pi_1^\prime (u,d,b)~, \\
\label{eZ23}
- \Pi_2 (u,d,b) \es  \Pi_1 (b,d,u) +  \Pi_1^\prime (u,b,d)~.
\eea
With the replacement $b \lrar u$ Eq. (\ref{eZ21}) goes to Eq. (\ref{eZ23})
and with the replacement $b \lrar d$ Eq. (\ref{eZ22}) goes to
Eq. (\ref{eZ23}), as the result of which we get
\bea
\label{eZ24}
 \Pi_1^\prime (u,b,d) =  \Pi_1 (b,d,u)~,~\mbox{\rm and}, \\
\label{eZ25}
 \Pi_1^\prime (d,b,u) =  \Pi_1 (b,d,u)~.
\eea

Using  these relations and Eq. (\ref{eZ19}), we get the following relation
for the invariant
function responsible for the $\Sigma_b^{\ast 0} \rar \Lambda_b \rho^0$
transition
\bea
\label{eZ26}
\sqrt{3} \Pi^{\Sigma_b^{\ast 0} \rar \Lambda_b \rho^0} \es 
- g_{\rho uu} \Big[ 
2\Pi_1 (u,b,d) + \Pi_1 (u,d,b) \Big] + g_{\rho dd} \Big[ 
2 \Pi_1 (d,b,u) + \Pi_2 (d,u,b) \Big] \nnb \\
\ar  g_{\rho bb} \Big[ \Pi_1(b,d,u) -    
\Pi_1 (b,u,d) \Big]~.
\eea

Comparing these results with Eq. (\ref{eZ15}), we finally get
\bea
\label{eZ27}
\widetilde{\Pi}_1 (u,d,b) \es -{1\over \sqrt{3}} 
\Big[2 \Pi_1 (u,b,d) + \Pi_1 (u,d,b) \Big]~, \nnb \\     
\widetilde{\Pi}_1^\prime (u,d,b) \es {1\over \sqrt{3}}
\Big[2 \Pi_1 (d,b,u) + \Pi_1 (d,u,b) \Big]~,\nnb \\
\widetilde{\Pi}_2 (u,d,b) \es {1\over \sqrt{3}}
\Big[\Pi_1 (b,d,u) - \Pi_1 (b,u,d) \Big]~.
\eea

Relations among invariant functions describing sextet--antitriplet
transitions involving light vector mesons are presented in the Appendix.

For obtaining sum rules for the coupling constants the expressions of the correlation
functions from QCD side are needed. The corresponding correlation functions can be evaluated 
in deep Euclidean region, $-p_1^2 \rar \infty$, $-p_2^2 \rar \infty$, as has
already been mentioned, using the OPE. 
In the Light Cone version of the QCD sum rules formalism, the OPE is performed with
respect to twists of the corresponding nonlocal operators. In this expansion
the DA's of the vector mesons appear as the main
nonperturbative parameters.
Up to twist--4 accuracy,
matrix elements $\lla V(q) \vel \bar{q}(x) \Gamma q(0) \ver 0 \rra$ and 
$\lla V(q) \vel \bar{q}(x) G_{\mu\nu} q(0) \ver 0 \rra$ are determined in
terms of the DA's of the vector mesons,
where $\Gamma$ represents the Dirac matrices relevant to the case under
consideration, and $G_{\mu\nu}$ is the gluon field strength tensor. 
The definitions of these DA's for vector mesons are presented in \cite{RZ17}-\cite{RZ18}.
Having the expressions of the heavy and light quark propagators ( see \cite{RZ19},\cite{RZ20} ) and 
the DA's for the light vector mesons we can straightforwardly 
calculate the correlation 
functions from the QCD side. Equating both representations of correlation
function and separating coefficients of Lorentz structures 
$\rlap/{\varepsilon}\rlap/{p}\gamma_5 q_\mu$,
$\rlap/{q} \rlap/{p} \gamma_5 (p\cdot\varepsilon) q_\mu$ and $\rlap/{q}
\rlap/{p} \gamma_5 \varepsilon_\mu$, and
applying Borel transformation to the variables $p^2$ and $(p+q)^2$ on both 
sides of the correlation functions, which suppresses the
contributions of the higher states and continuum, we obtain sum rules for the coupling
constants $g_1$, $g_2$ and $g_3$:
\bea
\label{eZ28}
g_k \es \kappa_k\Pi_1^{(k)}(M^2)~,
\eea
where
\bea
\label{eZ29}
\kappa_1 \es {\kappa_2\over m_1+m_2}~, \nnb \\ 
\nnb \\
\kappa_2 \es {1\over \lambda_{B_Q} \lambda_{B_Q^*}}e^{\left({{m_1^2 \over M_1^2} + 
{m_2^2 \over M_2^2} + {m_V^2 \over{M_1^2 + M_2^2}}}\right)}~, \nnb \\
\kappa_3 \es {\kappa_2\over m_V^2}~.   
\eea

As has already been noted, relations between invariant functions are independent of the Lorenz structures,
but their explicit expressions are  structure dependent. So we have introduced
extra upper index $k$ in brackets for each coupling, and $k$=1,2 and 3 corresponds  the choice of the Lorenz structures
$\rlap/{\varepsilon}\rlap/{p}\gamma_5 q_\mu$,
$\rlap/{q} \rlap/{p} \gamma_5 (p\cdot\varepsilon) q_\mu$ and $\rlap/{q}
\rlap/{p} \gamma_5 \varepsilon_\mu$, respectively.
In this equation $M_1^2 $ and $M_2^2$ are Borel parameters in the initial and final baryon channels.
Since the masses of the initial and final baryons are close to each other we put
$M_1^2=M_2^2=M^2$. Residues $\lambda_{B_Q^*}$ and $\lambda_{B_Q}$ of the heavy baryons of
spin--3/2 and --1/2 
have been calculated in \cite{RZ21}. As the explicit formulae for $\Pi_1^{(k)}$, $k$=1,2,3 are lengthy and not very instructive
we do not present it in the body of our article.
\section{Numerical analysis}
In this section, we present our numerical results on the strong coupling constants of the light vector mesons
with the sextet and antitriplet of heavy baryons. The main input parameters in LCSR for the coupling constants
are the DA's of the light vector mesons.
These DA's and its parameters are taken from \cite{RZ17} and \cite{RZ18}. 
The LCSR's also contain the following auxiliary parameters: Borel parameter $M^2$, threshold of the continuum $s_0$
and the parameter $\beta$ of the interpolating currents of the spin--1/2 baryons.
Obviously any physical quantity should be independent of these auxiliary parameters. Therefore, we should find
''working regions" of these parameters where coupling constants of the $B_Q^* B_Q V$ transitions are practically
independent of them. We proceed along the same scheme as is presented in \cite{RZ11}-\cite{RZ16}.
In order to find the ''working region" of $M^2$, we require that the continuum and higher state contributions
should be less then half of the dispersion integral while the contribution of the higher terms proportional to
$1/M^2$ be less then 25\% of the total result.These two requirements give the
''working region" of $M^2$
in the range 15 GeV$^2 \leq M^2 \leq$ 30 GeV$^2$ for baryons with the single $b$ quark and
4 GeV$^2 \leq M^2 \leq$ 8 GeV$^2$ for the charmed baryons, respectively. The continuum threshold is not totally arbitrary but is correlated to the energy of the
 first excited states with the same quantum numbers as the interpolating currents.
This parameter is chosen in the range 
($m_{B^*_Q}+0.5$ GeV)$^2 \leq s_0 \leq$ ($m_{B^*_Q}+0.7$ GeV)$^2 $. Our results show weak dependence on this parameter in this working region.

As an example let us consider the transition $\Xi^{*+}_c\rar \Xi^{\prime+}_c \rho^0$ and show in what way the
coupling constants $g_1$, $g_2$ and $g_3$ are determined.
In Figs. (1)--(3) we depict the dependence of the coupling constants $g_1$, $g_2$ and $g_3$ on  $M^2$
at $s_0=10.5$ GeV$^2$ and several different fixed values of $\beta $. It is seen that the coupling constants depend weakly 
on  $M^2$ in the ''working region".
Now, we proceed to calculate the working region of the general parameter $\beta$ entering the interpolating currents of the spin--1/2 particles. This parameter
 is also not physical quantity, hence we should look for an optimal working region at which the dependence of our results also on this parameter is weak. Due to the truncated OPE, in general,
 the zeros of the sum rules for strong coupling constants and residues are not coincide. These points and close to these points are an artifact
 of the using truncated OPE and hence the "working" region of $cos\theta$ should be far from these region. In Figs. (4)--(6) (as an example) we present the dependence of 
the coupling constants $g_1$, $g_2$ and $g_3$ of this transition on $\cos\theta$, where
$\tan\theta=\beta$,
at three fixed values of $s_0$ and at a fixed value of $M^2$. From these figures it is easily seen
that the coupling constants $g_1$, $g_2$ and $g_3$ are practically unchanged while $\cos\theta$
is varying in the domain -0.5$\leq \cos\theta \leq 0.3$ and  weakly depend on $s_0$.  Plotting all the considered strong coupling constants for all allowed transitions versus $\cos\theta$, we see that this working region is approximately
 common and  optimal one to achieve
reliable sum rules for all cases.
From our analysis we obtain $g_1^{\Xi^{*+}_c \to \Xi^{\prime+}_c \rho^0} = (2.6 \pm 0.5)~GeV^{-1}$,
$g_2^{\Xi^{*+}_c \to \Xi^{\prime+}_c \rho^0} =(0.9 \pm 0.2)~GeV^{-2}$, $g_3^{\Xi^{*+}_c \to \Xi^{\prime+}_c \rho^0}
=(22 \pm 4)~GeV^{-2}$.
The results for the coupling constants of other transitions are put into the Tables 3 and 4.
For completeness, in these Tables we also present results of the nonrelativistic quark model (NRQM) on these
couplings in terms of a constant $c$.
From these tables we can conclude that predictions of the general and the Ioffe currents are very close to each other.
We also see that ratios of the decays considered are also in good agreement with the predictions of the
the nonrelativistic quark model.
Finally it should be noted that some of the coupling constants related with $\rho^0$, $\omega$ and $\phi $  mesons
were studied in \cite{RZ22} with the Ioffe interpolating currents. 
The results obtained in that work do partially agree or disagree
compared to our predictions.

\section{Conclusion}

In the present work, we have studied the $B_Q^* B_Q V$ vertices within the LCSR method.
These vertices are parametrized with three coupling constants. We have calculated them for all 
the $B_Q^* B_Q V$ transitions with light vector mesons. The main result is that the correlation functions 
responsible for the coupling of the light vector mesons with the heavy sextet baryons of the
spin--3/2 and the heavy sextet and
antitriplet baryons of the spin--1/2 are described in terms of only one invariant function for
each Lorenz structure while the relations between the different transitions are structure independent.


\begin{table}[h]

\renewcommand{\arraystretch}{1.3}
\addtolength{\arraycolsep}{-0.5pt}
\small
$$
\begin{array}{|l|r@{\pm}l|r@{\pm}l|r@{\pm}l|r@{\pm}l|r@{\pm}l|r@{\pm}l|c|}
\hline \hline  
\mbox{\small{~~~\,transition}}
&    \multicolumn{2}{c}{\mbox{$g_1$}}  & \multicolumn{2}{c|}{\mbox{$g_1^{Ioffe}$}} 
&    \multicolumn{2}{c}{\mbox{$g_2$}}  & \multicolumn{2}{c|}{\mbox{$g_2^{Ioffe}$}} &
     \multicolumn{2}{c}{\mbox{$g_3$}}  & \multicolumn{2}{c|}{\mbox{$g_3^{Ioffe}$}} & 
     \mbox{\small{\,NRQM}} \\ \hline
\Sigma^{\ast 0}_b \rar \Sigma^-_b \rho^+               &  4.2&1.0  &  4.6&1.2  &  0.7&0.2   & 0.7&0.2   &  84&20  &  90&23   &(2/3)c\\ 
\Sigma^{\ast +}_b \rar \Sigma^+_b \omega               &  3.8&1.1  &  4.2&1.1  &  0.6&0.2   & 0.7&0.2   &  70&20  &  74&21   &(2/3)c\\ 
\Sigma^{\ast -}_b \rar \Lambda_b \rho^-                &  8.0&2.1  &  8.4&1.4  &  0.7&0.2   & 0.8&0.2   & 148&38  & 154&26   &(2/\sqrt3)c\\ 
\Sigma^{\ast +}_b\rar\Xi^{\prime 0}_b K^{\ast +}       &  4.6&1.4  &  5.2&1.3  &  0.9&0.3   & 1.2&0.4   &  64&18  &  70&18   &(2/3)c\\
\Sigma^{\ast 0}_b \rar \Xi^0_b K^{\ast 0}              &  6.4&1.7  &  6.8&1.2  &  0.9&0.2   & 1.1&0.2   &  87&20  &  88&15   &(\sqrt{2/3})c\\
\Xi^{\ast 0}_b\rar\Xi^{\prime 0}_b \rho^0              &  2.4&0.7  &  2.7&0.7  &  0.4&0.1   & 0.4&0.1   &  48&10  &  52&14   &(1/3)c\\
\Xi^{\ast 0}_b\rar\Xi^{\prime 0}_b \omega              &  2.4&0.5  &  2.4&0.6  &  0.4&0.1   & 0.3&0.1   &  40&10  &  42&12   &(1/3)c\\
\Xi^{\ast 0}_b\rar\Xi^{\prime 0}_b \phi                &  3.2&1.0  &  3.6&1.0  &  0.3&0.1   & 0.4&0.1   &  35&11  &  39&12   &(\sqrt2/3)c\\
\Xi^{\ast 0}_b\rar \Sigma^+_b K^{\ast -}               &  4.7&1.4  &  4.9&1.4  &  1.0&0.3   & 1.2&0.4   &  64&16  &  66&16   &(2/3)c\\
\Xi^{\ast 0}_b\rar \Omega^-_b K^{\ast +}               &  5.3&1.6  &  5.9&1.7  &  1.0&0.2   & 1.3&0.3   &  74&21  &  83&20   &(2/3)c\\
\Xi^{\ast 0}_b\rar\Xi^0_b\rho^0                        &  4.3&1.1  &  4.4&0.8  &  0.4&0.1   & 0.4&0.1   &  82&20  &  84&16   &(1/\sqrt3)c\\
\Xi^{\ast 0}_b\rar\Xi^0_b \omega                       &  4.3&1.1  &  4.4&0.8  &  0.4&0.1   & 0.4&0.1   &  82&20  &  84&16   &(1/\sqrt3)c\\
\Xi^{\ast 0}_b\rar\Xi^0_b \phi                         &  5.9&1.6  &  6.4&1.1  &  0.7&0.2   & 0.7&0.2   &  65&17  &  69&12   &(\sqrt{2/3})c\\
\Xi^{\ast 0}_b\rar\Lambda_b \bar{K}^{\ast 0}           &  6.3&1.7  &  6.6&1.0  &  0.5&0.1   & 0.6&0.2   &  86&20  &  88&15   &(\sqrt{2/3})c\\
\Omega^{\ast -}_b\rar \Omega^-_b \phi                  &  7.2&1.8  &  8.2&2.2  &  1.0&0.2   & 1.2&0.3   &  85&15  &  90&25   &(2\sqrt2/3)c\\
\Omega^{\ast -}_b\rar\Xi^{\prime 0}_b K^{\ast -}       &  5.3&1.5  &  5.8&1.6  &  1.0&0.3   & 1.2&0.4   &  74&18  &  77&15   &(2/3)c\\
\Omega^{\ast -}_b \rar \Xi^-_b \bar{K}^{\ast 0}        &  9.7&2.5  & 10.0&2.0  &  1.3&0.3   & 1.5&0.3   & 136&31  & 133&25   &(2/\sqrt3)c\\
 \hline \hline
\end{array}
$$
\caption{The absolute values of the coupling constants $g_{1,2,3}$ for transitions of b-baryons. The couplings, $g_1$, $g_2$ and $g_3$ are in $GeV^{-1}$,   $GeV^{-2}$ and  $GeV^{-2}$, respectively.}
\renewcommand{\arraystretch}{1}
\addtolength{\arraycolsep}{-1.0pt}

\end{table}

\begin{table}[h]

\renewcommand{\arraystretch}{1.3}
\addtolength{\arraycolsep}{-0.5pt}
\small
$$
\begin{array}{|l|r@{\pm}l|r@{\pm}l|r@{\pm}l|r@{\pm}l|r@{\pm}l|r@{\pm}l|c|}
\hline \hline  
\mbox{\small{~~~~\,`transition}}
&    \multicolumn{2}{c}{\mbox{$g_1$}}  & \multicolumn{2}{c|}{\mbox{$g_1^{Ioffe}$}} 
&    \multicolumn{2}{c}{\mbox{$g_2$}}  & \multicolumn{2}{c|}{\mbox{$g_2^{Ioffe}$}} &
     \multicolumn{2}{c}{\mbox{$g_3$}}  & \multicolumn{2}{c|}{\mbox{$g_3^{Ioffe}$}} & 
     \mbox{\small{\,NRQM}} \\ \hline
\Sigma^{\ast +}_c \rar \Sigma^0_c \rho^+               &  5.0&1.0  &  5.7&0.5  &  1.0&0.2   & 1.2&0.2   &  40&7   &  45&5    &(2/3)c\\ 
\Sigma^{\ast ++}_c \rar \Sigma^{++}_c \omega           &  4.4&0.8  &  5.0&0.5  &  0.9&0.2   & 1.1&0.2   &  35&6   &  39&6    &(2/3)c\\ 
\Sigma^{\ast 0}_c\rar \Lambda_c^+ \rho^-               &  9.6&1.8  & 10.6&1.8  &  1.5&0.6   & 1.9&0.3   &  75&13  &  79&10   &(2/\sqrt3)c\\ 
\Sigma^{\ast ++}_c\rar\Xi^{\prime +}_c K^{\ast +}      &  5.4&1.0  &  6.0&0.5  &  1.5&0.5   & 2.3&0.4   &  29&4   &  32&3    &(2/3)c\\
\Sigma^{\ast +}_c \rar \Xi^+_c K^{\ast 0}              &  7.5&1.4  &  8.2&1.3  &  2.2&0.8   & 2.6&0.4   &  41&6   &  42&4    &(\sqrt{2/3})c\\
\Xi^{\ast +}_c\rar\Xi^{\prime +}_c \rho^0              &  2.6&0.5  &  3.0&0.3  &  0.9&0.2   & 1.0&0.2   &  22&4   &  25&4    &(1/3)c\\
\Xi^{\ast +}_c\rar\Xi^{\prime +}_c \omega              &  2.3&0.4  &  2.7&0.4  &  0.8&0.2   & 0.9&0.2   &  19&3   &  21&2    &(1/3)c\\
\Xi^{\ast +}_c\rar\Xi^{\prime +}_c \phi                &  3.6&0.7  &  4.1&0.6  &  0.9&0.2   & 1.1&0.2   &  17&3   &  20&2    &(\sqrt2/3)c\\
\Xi^{\ast +}_c\rar \Sigma^{++}_c K^{\ast -}            &  5.5&1.0  &  6.1&0.8  &  1.8&0.4   & 2.3&0.4   &  29&4   &  31&5    &(2/3)c\\
\Xi^{\ast +}_c\rar \Omega^0_c K^{\ast +}               &  5.6&1.0  &  6.5&1.0  &  2.0&0.5   & 2.4&0.5   &  32&5   &  35&5    &(2/3)c\\
\Xi^{\ast +}_c\rar\Xi^+_c\rho^0                        &  5.0&0.9  &  5.4&0.3  &  0.9&0.2   & 1.0&0.2   &  41&7   &  44&7    &(1/\sqrt3)c\\
\Xi^{\ast +}_c\rar\Xi^+_c \omega                       &  5.0&0.9  &  4.8&0.6  &  0.9&0.2   & 1.0&0.2   &  41&7   &  44&7    &(1/\sqrt3)c\\
\Xi^{\ast +}_c\rar\Xi^+_c \phi                         &  6.8&1.3  &  7.6&0.7  &  1.3&0.3   & 1.6&0.3   &  32&6   &  34&5    &(\sqrt{2/3})c\\
\Xi^{\ast +}_c\rar\Lambda^+_c \bar{K}^{\ast 0}         &  7.4&1.3  &  8.0&0.8  &  1.7&0.5   & 2.0&0.4   &  40&5   &  41&6    &(\sqrt{2/3})c\\
\Omega^{\ast 0}_c\rar \Omega^0_c \phi                  &  7.8&1.0  &  8.7&1.0  &  2.1&0.3   & 2.4&0.3   &  37&7   &  42&4    &(2\sqrt2/3)c\\
\Omega^{\ast 0}_c\rar\Xi^{\prime +}_c K^{\ast -}       &  5.8&1.0  &  6.4&0.8  &  2.1&0.4   & 2.4&0.3   &  32&4   &  34&3    &(2/3)c\\
\Omega^{\ast 0}_c \rar \Xi^0_c \bar{K}^{\ast 0}        & 11.0&2.0  & 12.0&2.0  &  3.5&0.6   & 4.4&0.6   &  63&9   &  65&7    &(2/\sqrt3)c\\
 \hline \hline
\end{array}
$$
\caption{The absolute values of the coupling constants $g_{1,2,3}$ for transitions of charmed baryons. The couplings, $g_1$, $g_2$ and $g_3$ are in $GeV^{-1}$,   $GeV^{-2}$ and  $GeV^{-2}$, respectively.}
\renewcommand{\arraystretch}{1}
\addtolength{\arraycolsep}{-1.0pt}

\end{table}


\newpage

\bAPP{}{}

Here in this appendix we present the expressions of the correlation functions
in terms of invariant function $\Pi_1$ involving 
$\rho$, $K^*$, $\omega$ and $\phi$ mesons.

\begin{itemize}
\item Correlation functions responsible for the 
sextet--sextet transitions.
\end{itemize}
\baeeq
\label{nolabel}
\Pi^{\Sigma_b^{\ast 0} \rar \Sigma_b^0     \rho^0 } \es 
{1\over \sqrt{2}} \Big[\Pi_1(u,d,b) - \Pi_1(d,u,b) \Big]~, \nnb \\
\Pi^{\Sigma_b^{\ast +} \rar \Sigma_b^+     \rho^0 } \es
\sqrt{2} \Pi_1(u,u,b)~, \nnb \\
\Pi^{\Sigma_b^{\ast -} \rar \Sigma_b^-     \rho^0 } \es
- \sqrt{2} \Pi_1(d,d,b)~, \nnb \\
\Pi^{\Xi_b^{\ast 0} \rar \Xi_b^{'0}     \rho^0 } \es 
{1\over \sqrt{2}} \Pi_1(u,s,b)~, \nnb \\
\Pi^{\Xi_b^{\ast -} \rar \Xi_b^{'-}     \rho^0 } \es 
- {1\over \sqrt{2}} \Pi_1(d,s,b)~, \nnb \\
\Pi^{\Sigma_b^{\ast +} \rar \Sigma_b^0     \rho^+ } \es 
\sqrt{2} \Pi_1(d,u,b)~, \nnb \\
\Pi^{\Sigma_b^{\ast 0} \rar \Sigma_b^-     \rho^+ } \es 
\sqrt{2} \Pi_1(u,d,b)~, \nnb \\
\Pi^{\Xi_b^{\ast 0} \rar \Xi_b^{'-}     \rho^+ } \es 
\Pi_1(d,s,b)~, \nnb \\
\Pi^{\Sigma_b^{\ast 0} \rar \Sigma_b^+     \rho^- } \es
\sqrt{2} \Pi_1(d,u,b)~, \nnb \\
\Pi^{\Sigma_b^{\ast -} \rar \Sigma_b^0     \rho^- } \es 
\sqrt{2} \Pi_1(u,d,b)~, \nnb \\
\Pi^{\Xi_b^{\ast -} \rar \Xi_b^{'0}     \rho^- } \es 
\Pi_1(u,s,b)~, \nnb \\
\Pi^{\Xi_b^{\ast 0} \rar \Sigma_b^+     K^{\ast-}} \es 
\sqrt{2} \Pi_1(u,u,b)~, \nnb \\
\Pi^{\Xi_b^{\ast -} \rar \Sigma_b^0     K^{\ast-}} \es 
\Pi_1(u,d,b)~, \nnb \\
\Pi^{\Omega_b^{\ast -} \rar \Xi_b^{'0}     K^{\ast-}} \es 
\sqrt{2} \Pi_1(s,s,b)~, \nnb \\
\Pi^{\Sigma_b^{\ast +} \rar \Xi_b^{'0}     K^{\ast+}} \es 
\sqrt{2} \Pi_1(u,u,b)~, \nnb \\
\Pi^{\Sigma_b^{\ast 0} \rar \Xi_b^{'-}     K^{\ast+}} \es 
\Pi_1(u,d,b)~, \nnb \\
\Pi^{\Xi_b^{\ast 0} \rar \Omega_b^-      K^{\ast+}} \es 
\sqrt{2}\Pi_1(s,s,b)~, \nnb \\
\Pi^{\Xi_b^{\ast 0} \rar \Sigma_b^0     \bar{K}^{\ast0}} \es 
\Pi_1(d,u,b)~, \nnb \\
\Pi^{\Xi_b^{\ast -} \rar \Sigma_b^-     \bar{K}^{\ast0}} \es 
\sqrt{2} \Pi_1(d,d,b)~, \nnb \\
\Pi^{\Omega_b^{\ast -}  \rar \Xi_b^{'-}    \bar{K}^{\ast0}} \es 
\sqrt{2} \Pi_1(s,s,b)~, \nnb \\
\Pi^{\Sigma_b^{\ast 0} \rar \Xi_b^{'0}     K^{\ast0}} \es 
\Pi_1(d,u,b)~, \nnb \\
\Pi^{\Sigma_b^{\ast -} \rar \Xi_b^{'-}     K^{\ast0}} \es 
\sqrt{2} \Pi_1(d,d,b)~, \nnb \\
\Pi^{\Xi_b^{\ast -} \rar \Omega_b^-     K^{\ast0}} \es 
\sqrt{2} \Pi_1(s,s,b)~, \nnb \\
\Pi^{\Sigma_b^{\ast 0} \rar \Sigma_b^0    \omega} \es 
{1\over \sqrt{2}}\Big[ \Pi_1(u,d,b) + \Pi_1(d,u,b)\Big]~, \nnb \\
\Pi^{\Sigma_b^{\ast +} \rar \Sigma_b^+    \omega} \es
\sqrt{2} \Pi_1(u,u,b)~, \nnb \\
\Pi^{\Sigma_b^{\ast -} \rar \Sigma_b^-    \omega} \es 
\sqrt{2} \Pi_1(d,d,b)~, \nnb \\
\Pi^{\Xi_b^{\ast 0} \rar \Xi_b^{'0}     \omega } \es
{1\over \sqrt{2}} \Pi_1(u,s,b)~, \nnb \\
\Pi^{\Xi_b^{\ast -} \rar \Xi_b^{'-}     \omega } \es
{1\over \sqrt{2}} \Pi_1(d,s,b)~, \nnb \\
\Pi^{\Xi_b^{\ast 0} \rar \Xi_b^{'0}     \phi } \es
\Pi_1(s,u,b)~, \nnb \\
\Pi^{\Xi_b^{\ast -} \rar \Xi_b^{'-}     \phi } \es
\Pi_1(s,d,b)~, \nnb \\
\Pi^{\Omega_b^{\ast -} \rar \Omega_b^-     \phi } \es
2 \Pi_1(s,s,b)~. \nnb
\eaeeq

\begin{itemize}
\item Correlation functions responsible for the 
sextet--antitriplet transitions.
\end{itemize}
\baeeq
\label{nolabel}
\Pi^{\Xi_b^{\ast 0} \rar \Xi_b^0     \rho^0 } \es
{1\over \sqrt{2}} \widetilde{\Pi}_1(u,s,b)~, \nnb \\
\Pi^{\Xi_b^{\ast -} \rar \Xi_b^-     \rho^0 } \es
{1\over \sqrt{2}} \widetilde{\Pi}_1(d,s,b)~, \nnb \\
\Pi^{\Sigma_b^{\ast 0} \rar \Lambda_b     \rho^0 } \es
{1\over \sqrt{2}} \Big[\widetilde{\Pi}_1(u,d,b) + \widetilde{\Pi}_1(d,u,b) \Big]~, \nnb \\
\Pi^{\Sigma_b^{\ast -} \rar \Lambda_b     \rho^- } \es
\sqrt{2} \widetilde{\Pi}_1(u,d,b)~, \nnb \\
\Pi^{\Xi_b^{\ast -} \rar \Xi_b^-      \rho^- } \es
\widetilde{\Pi}_1(d,s,b)~, \nnb \\
\Pi^{\Sigma_b^{\ast +} \rar \Lambda_b     \rho^+ } \es
- \sqrt{2} \widetilde{\Pi}_1(d,u,b)~, \nnb \\
\Pi^{\Xi_b^{\ast 0} \rar \Xi_b^-      \rho^+ } \es
\widetilde{\Pi}_1(u,s,b)~, \nnb \\
\Pi^{\Omega_b^{\ast -} \rar \Xi_b^-      \bar{K}^{\ast0}} \es
- \sqrt{2} \widetilde{\Pi}_1(s,s,b)~, \nnb \\ 
\Pi^{\Xi_b^{\ast 0} \rar \Lambda_b    \bar{K}^{\ast0}} \es
- \widetilde{\Pi}_1(d,u,b)~, \nnb \\
\Pi^{\Sigma_b^{\ast 0} \rar \Xi_b^0           K^{\ast0}} \es
- \widetilde{\Pi}_1(d,u,b)~, \nnb \\
\Pi^{\Sigma_b^{\ast -} \rar \Xi_b^-           K^{\ast0}} \es
\sqrt{2} \widetilde{\Pi}_1(d,d,b)~, \nnb \\
\Pi^{\Sigma_b^{\ast 0} \rar \Xi_b^-           K^{\ast+}} \es
\widetilde{\Pi}_1(u,d,b)~, \nnb \\
\Pi^{\Omega_b^{\ast -} \rar \Xi_b^0           K^{\ast-}} \es
\sqrt{2} \widetilde{\Pi}_1(s,s,b)~, \nnb \\ 
\Pi^{\Xi_b^{\ast 0} \rar \Xi_b^0          \omega}     \es
{1\over \sqrt{2}} \widetilde{\Pi}_1(u,s,b)~, \nnb \\
\Pi^{\Xi_b^{\ast -} \rar \Xi_b^-          \omega}     \es    
{1\over \sqrt{2}} \widetilde{\Pi}_1(d,s,b)~, \nnb \\    
\Pi^{\Sigma_b^{\ast 0} \rar \Lambda_b        \omega}     \es
{1\over \sqrt{2}} \Big[\widetilde{\Pi}_1(u,d,b) - \widetilde{\Pi}_1(d,u,b) \Big]~, \nnb \\
\Pi^{\Xi_b^{\ast 0} \rar \Xi_b^0          \phi}       \es
- \widetilde{\Pi}_1(s,u,b)~, \nnb \\
\Pi^{\Xi_b^{\ast -} \rar \Xi_b^-          \phi}       \es
- \widetilde{\Pi}_1(s,d,b)~. \nnb
\eaeeq

The expressions for the charmed baryons can easily be obtained by making the
replacement $b \rar c$ and adding to charge of each baryon a positive unit
charge. 

\eAPP

\begin{figure}[t]
\begin{center}
\scalebox{0.69}{\includegraphics{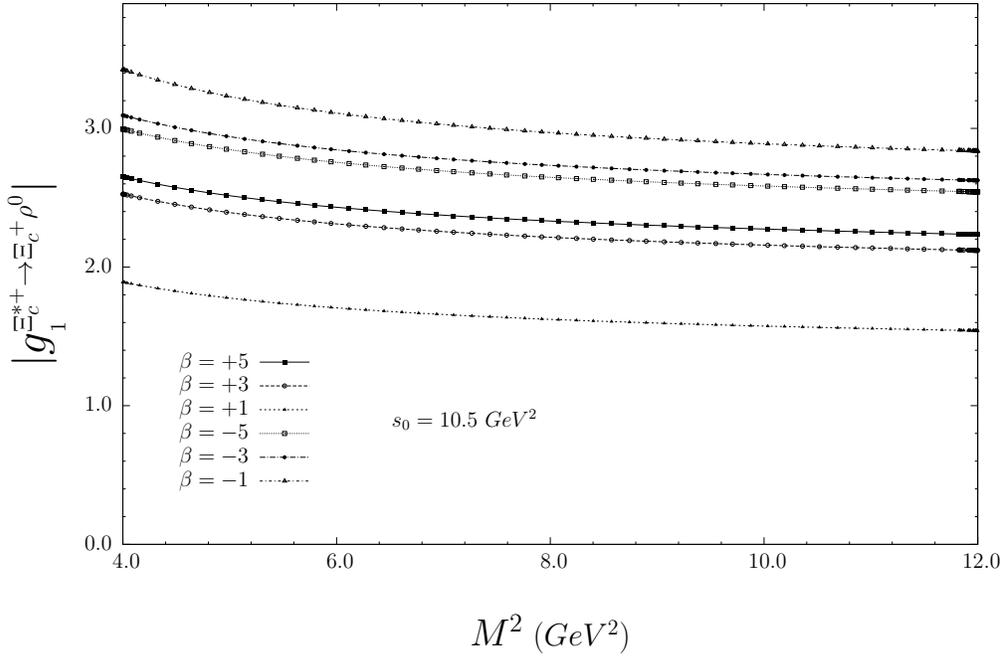}}
\end{center}
\caption{The dependence of the strong coupling constant $g_1$ for
the $\Xi_c^{*+} \rar \Xi_c^{'+} \rho^0$ transition on the Borel mass
parameter $M^2$ at several
different fixed values of $\beta$, and at $s_0=10.5~GeV^2$.}
\end{figure}

\begin{figure}[b]
\begin{center}
\scalebox{0.69}{\includegraphics{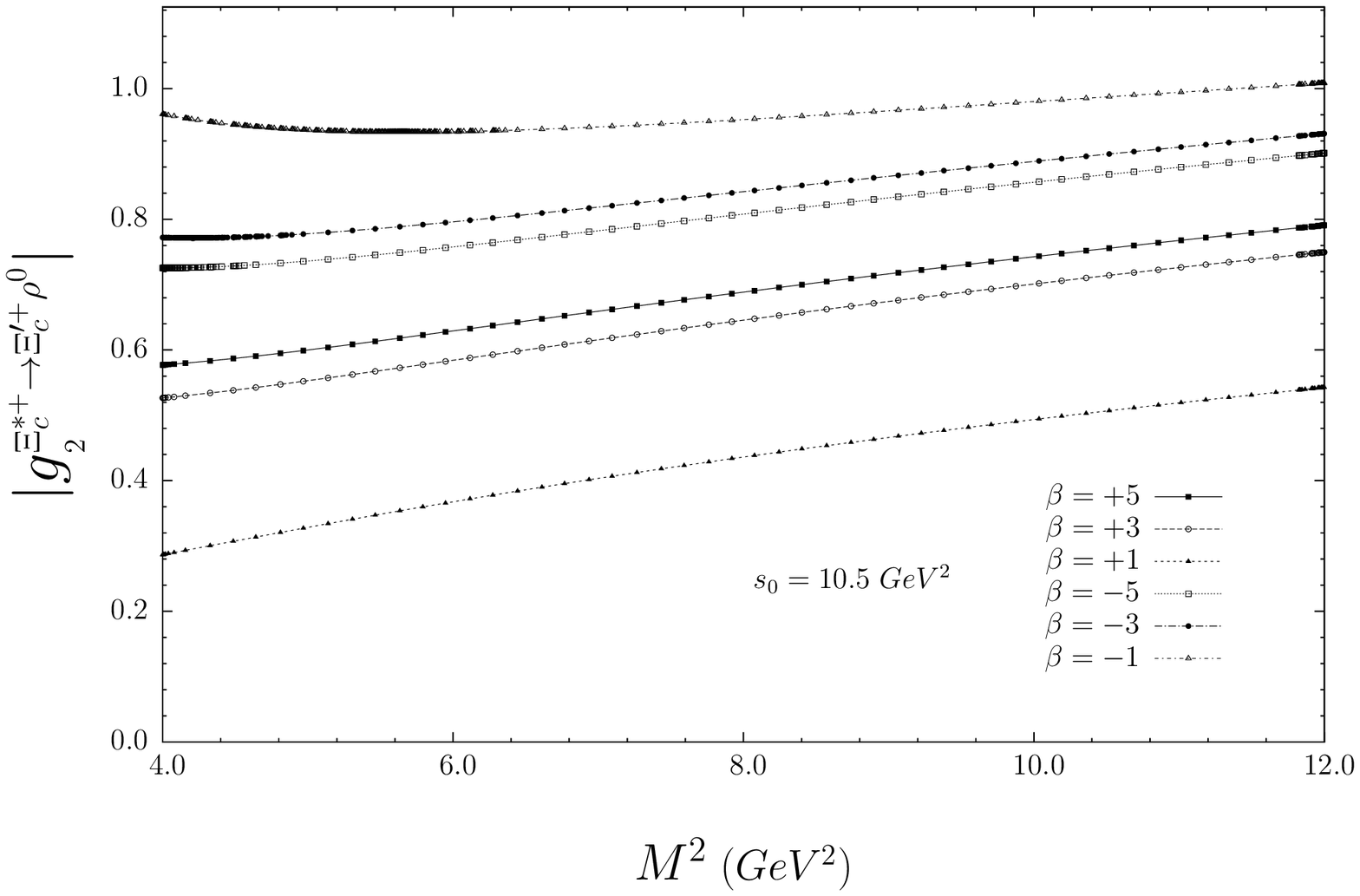}}
\end{center}
\caption{The same as Fig. (1), but for the strong coupling constant
$g_2$.}
\end{figure}

\begin{figure}[t]
\begin{center}
\scalebox{0.69}{\includegraphics{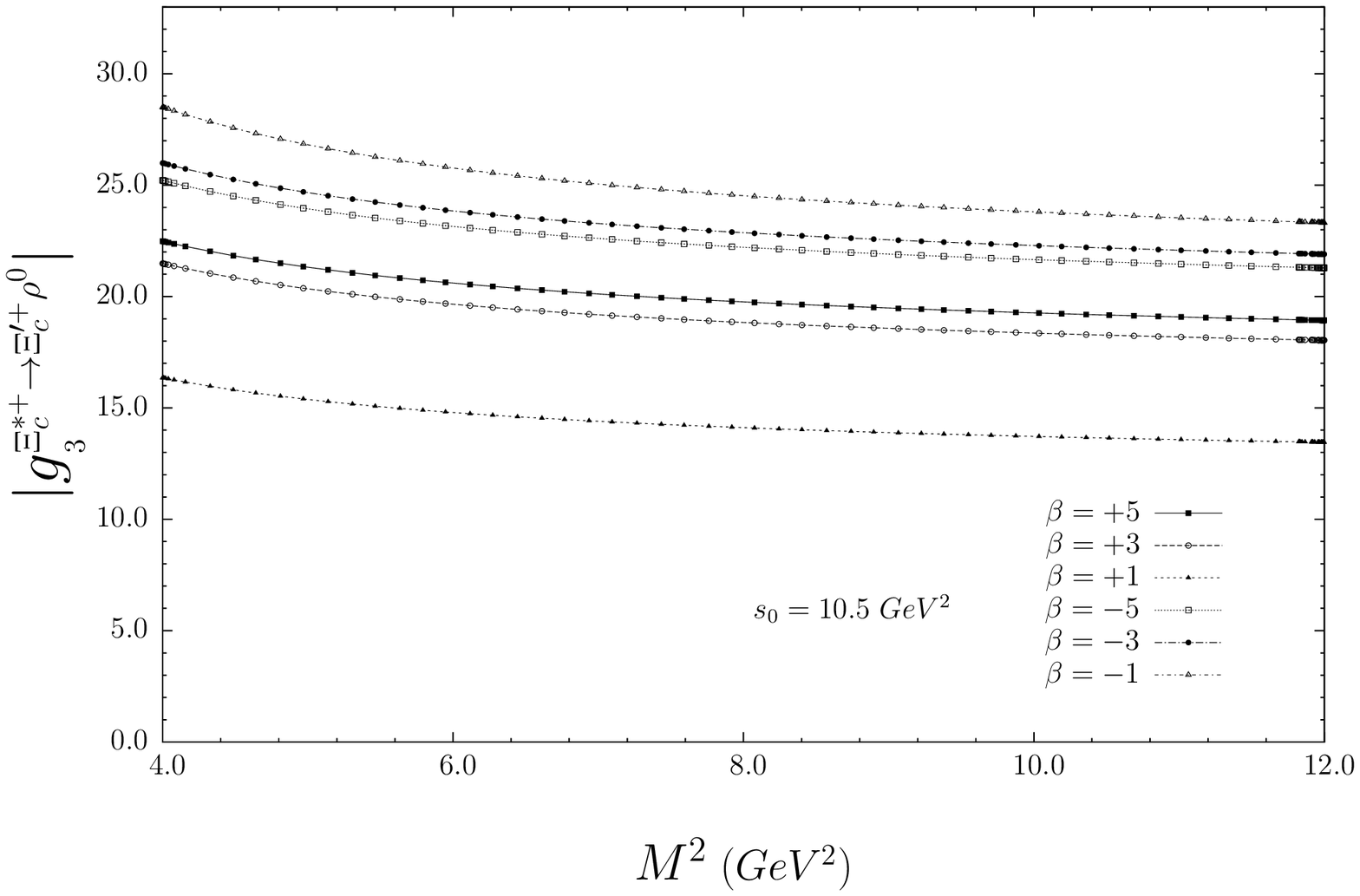}}
\end{center}
\caption{ The same as Fig. (1), but for the strong coupling constant   
$g_3$.}
\end{figure}

\begin{figure}[b]
\begin{center}
\scalebox{0.69}{\includegraphics{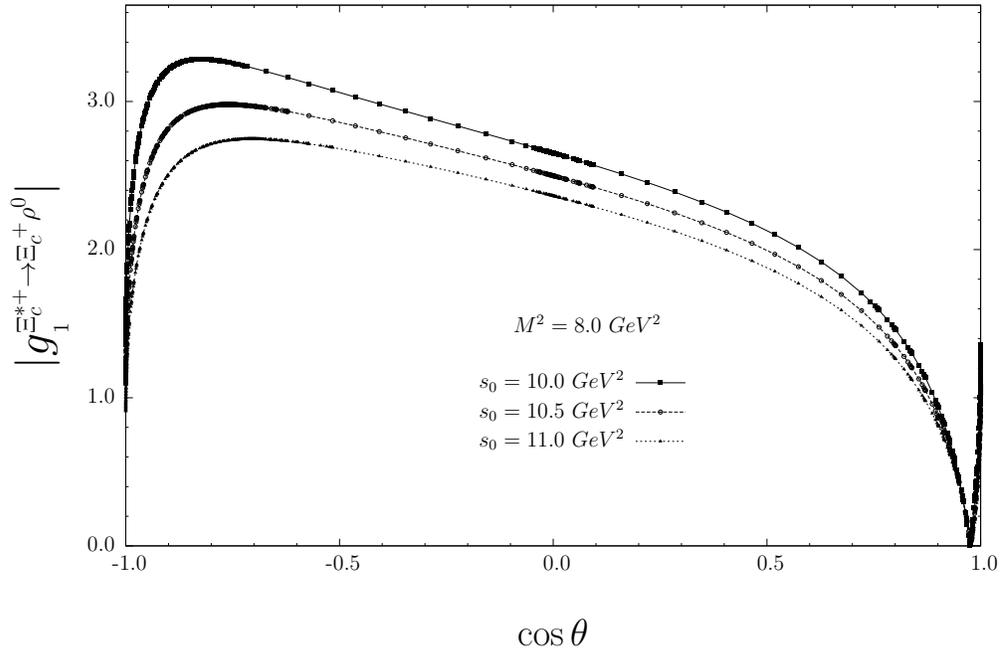}}
\end{center}
\caption{The dependence of the strong coupling constant $g_1$ for
the $\Xi_c^{*+} \rar \Xi_c^{'+} \rho^0$ transition on $\cos\theta$
at several different fixed values of $s_0$, and at $M^2= 8.0~GeV^2$.} 
\end{figure}

\begin{figure}[t]
\begin{center}
\scalebox{0.69}{\includegraphics{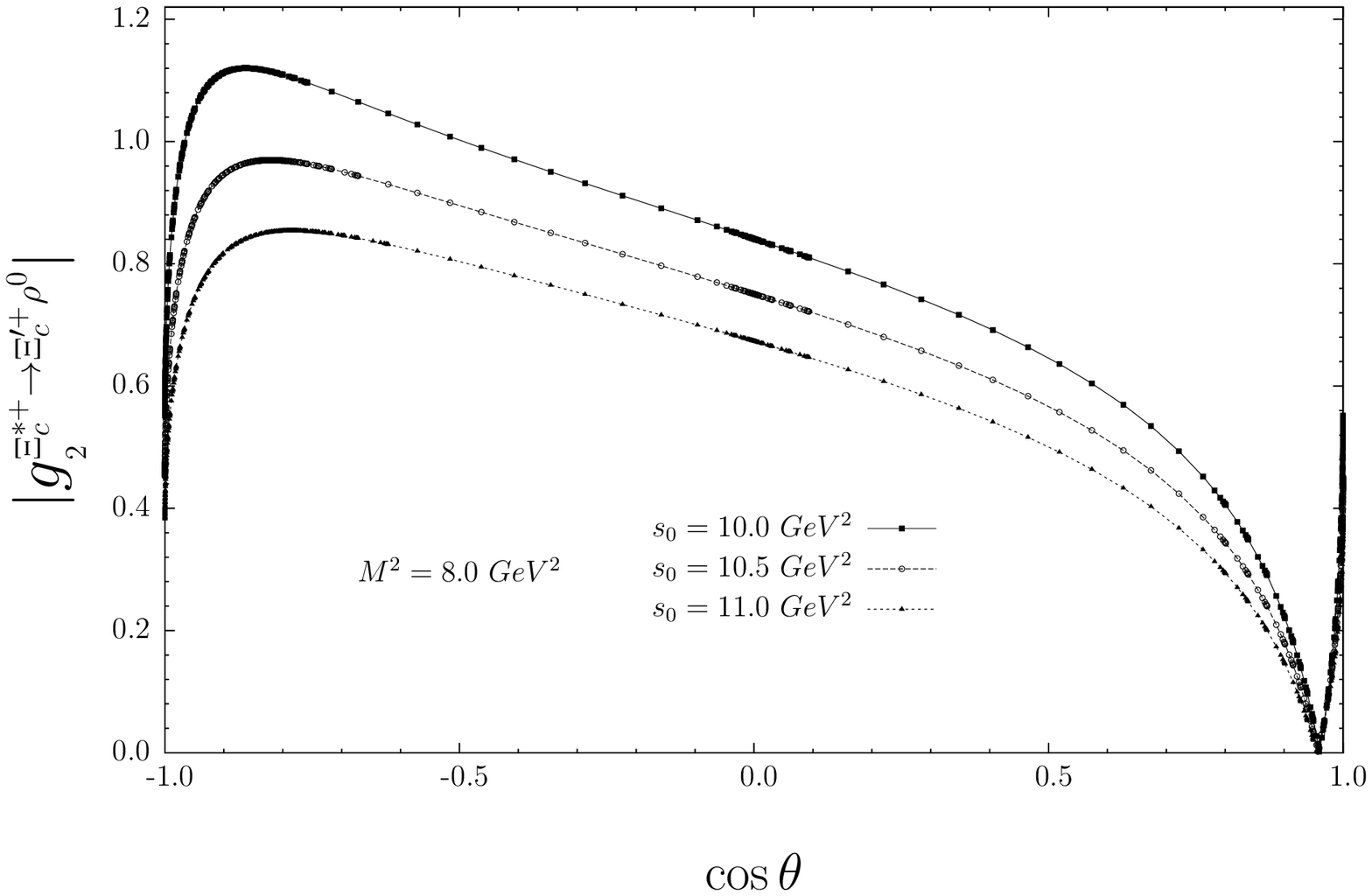}}
\end{center}
\caption{The same as Fig. (4), but for the strong coupling constant   
$g_2$.}
\end{figure}

\begin{figure}[b]
\begin{center}
\scalebox{0.69}{\includegraphics{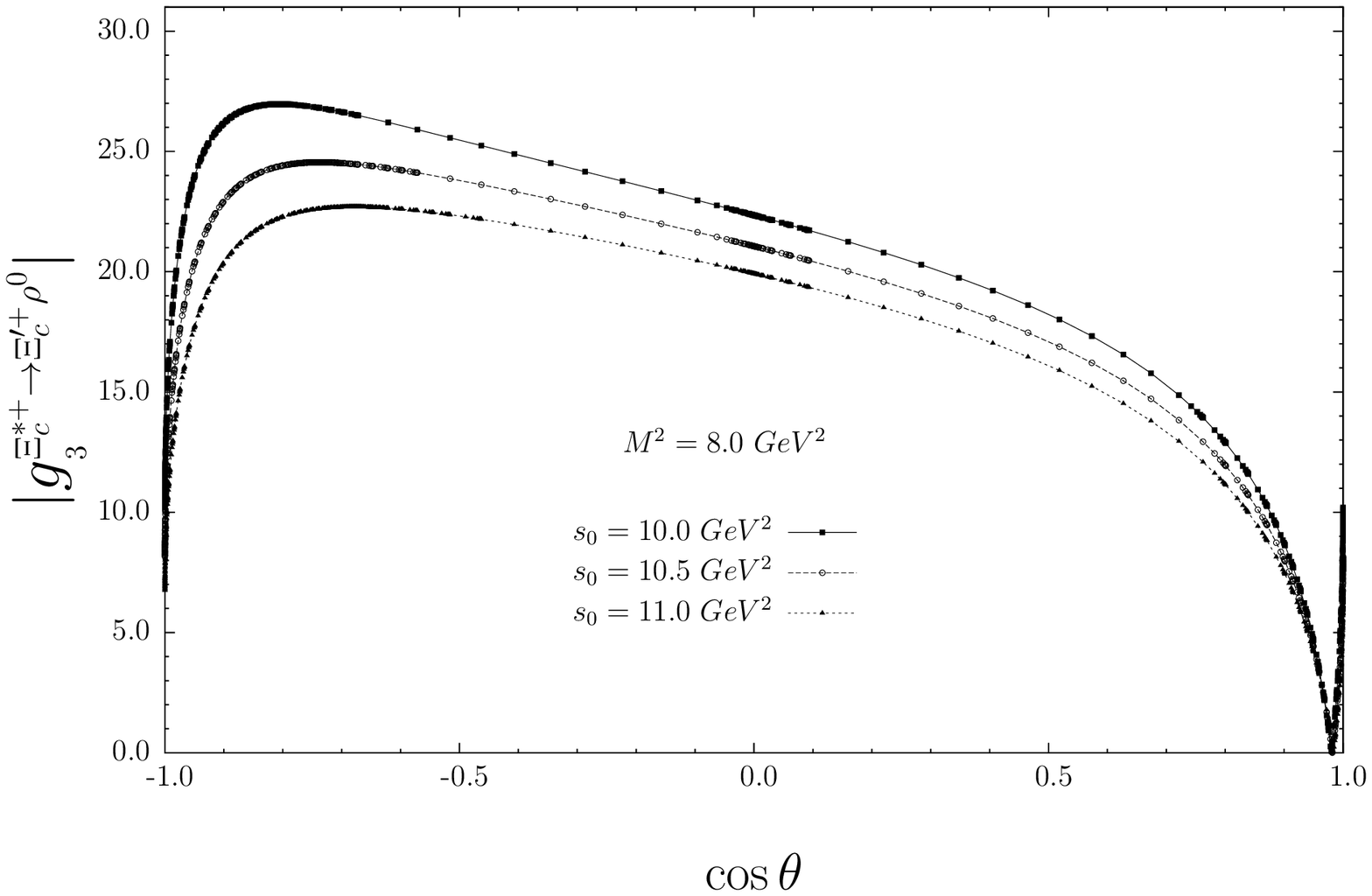}}
\end{center}
\caption{The same as Fig. (4), but for the strong coupling constant   
$g_3$. }
\end{figure}

\end{document}